\documentclass[12pt]{article}
\pdfoutput=1
\usepackage{jheppub}

\newcommand{\EQ}[1]{\begin{align}\begin{split} #1 
\end{split}\end{align}}
\newcommand{\eq}[1]{\begin{equation}{ #1 
}\end{equation}}
\def\BR{\mathbb R}\def\IZ{\mathbb{Z}}
\def\CA{{\cal A}}\def\CG{{\cal G}}\def\CN{{\cal N}}
\def\CR{{\cal R}}\def\CW{{\cal W}}\def\CY{{\cal Y}}\def\CZ{{\cal Z}}
\def\a{\alpha}\def\b{\beta}\def\g{\gamma}\def\e{\epsilon}\def\z{\zeta}
\def\th{\theta}\def\l{\lambda}\def\m{\mu}\def\t{\tau}\def\w{\omega}

\begin{document}
\title{\Large A 5d/3d duality from relativistic integrable system}
\author[]{Heng-Yu Chen${}^{1,2,3}$, Timothy J. Hollowood${}^{4}$ and Peng Zhao${}^{5}$}
\affiliation{$^1$Department of Physics and Center for Theoretical Sciences, \\
National Taiwan University, Taipei 10617, Taiwan}
\affiliation{$^2$Department of Physics, University of Cincinnati, Cincinnati OH 45221, USA}
\affiliation{$^3$Department of Physics and Astronomy, University of Kentucky, Lexington KY 40506, USA}
\affiliation{$^4$Department of Physics, Swansea University, Swansea SA2 8PP, UK}
\affiliation{$^5$DAMTP, University of Cambridge, Cambridge CB3 0WA, UK}
\emailAdd{heng.yu.chen@phys.ntu.edu.tw}
\emailAdd{t.hollowood@swansea.ac.uk}
\emailAdd{p.zhao@damtp.cam.ac.uk}
\abstract
{We propose and prove a new exact duality between the F-terms of supersymmetric gauge theories in five and three dimensions with adjoint matter fields. The theories are compactified on a circle and are subject to the $\Omega$ deformation. In the limit proposed by Nekrasov and Shatashvili, the supersymmetric vacua become isolated and are identified with the eigenstates of a quantum integrable system. The effective twisted superpotentials are the Yang-Yang functional of the relativistic elliptic Calogero-Moser model. We show that they match on-shell by deriving the Bethe ansatz equation from the saddle point of the five-dimensional partition function. We also show that the Chern-Simons terms match and extend our proposal to the elliptic quiver generalizations.}

\begin{flushright} DAMTP-2012-38 \end{flushright}
\maketitle
\section{Introduction}

The intimate connection between supersymmetric gauge theories and integrable models has been an extremely fertile research theme over the past few decades (see \cite{Gorsky:2000px} for a review). One of the best-known elements of this story is that the Seiberg-Witten curves of the four-dimensional ${\mathcal N}=2$ gauge theories can be identified with the spectral curves of classical integrable systems. Recently, a striking proposal of Nekrasov and Shatashvili extends this correspondence to \emph{quantum} integrable models by subjecting the gauge theories to a special limit of the so-called $\Omega$-deformation \cite{Nekrasov2009A}. This latter deformation introduces
chemical potentials $(\e_{1}, \e_{2})$ associated to the rotations in two orthogonal planes in ${\mathbb R}^4$. 
The Nekrasov-Shatashvili limit corresponds to taking $(\e_{1}, \e_{2}) \to (\e, 0)$ where one of the rotation parameter is turned off resulting in an $\CN = (2,2)$ supersymmetry being preserved in the fixed plane.\footnote{The string theory realization of the $\Omega$-background has recently been studied in \cite{Reffert:2011dp, Hellerman:2011mv, Hellerman:2012zf}.} In the deformed theory, $\e$ plays the role of Planck's constant $\hbar$ and the resulting geometry can be interpreted as a kind of quantization of the Seiberg-Witten curve. 

The correspondence can be naturally extended to gauge theories with eight supercharges in five dimensions compactified on a circle. These are known to lead to effective four-dimensional theories whose Seiberg-Witten curve corresponds to a relativistic integrable system where the radius $R$ of the compactification circle plays the role of the inverse speed of light \cite{Nekrasov:1996cz}. In the four-dimensional limit as $R \to 0$, we recover the non-relativistic version of the integrable system. 

The low energy physics of the four-dimensional theory is described by an effective twisted superpotential $\CW^\text{4d}$, a multi-valued function on the Coulomb branch defined in terms of Nekrasov's instanton partition function by 
\eq{
\CW^\text{4d}(\vec a, \e) = \lim_{(\e_{1},\e_{2}) \to (\e, 0)}\e_{2}\log \CZ^\text{4d} (\vec a, \e_{1},\e_{2}) - 2\pi i~\!\vec n \cdot \vec a\ ,
}
where the branch choice $\vec n \in {\mathbb{Z}}^{L}$ parametrizes the supersymmetric vacua and
the quantity $\e_1\e_{2}\log \CZ^\text{4d} (\vec a, \e_{1},\e_{2})$ in the Nekrasov-Shatashvili limit ${(\e_{1},\e_{2}) \to (\e, 0)}$ is the analogue of the prepotential in the deformed background. The freedom in $\vec n$ is also related to turning on the two-dimensional theta angle, which is equivalent to quantized electric flux via the Coleman effect. The vacuum equation is given by
\eq{
\frac{\partial\CW^\text{4d}}{\partial \vec a} = 0\ ,\label{typeA}
}
implying that the dual variable 
$\vec a^{D}=\vec n\epsilon$ is quantized in units of $\e$. 
Equivalently, we can go to a different electromagnetic duality frame via the Legendre transform
\eq{
\CW^{D}(\vec a^{D},\e) = \CW^\text{4d}(\vec a, \e) - \frac{2\pi i}{\e}\vec a\cdot \vec a^{D},
\label{legendre}}
which correspond to turning on quantized units of magnetic flux.
The vacuum equation is now
\eq{
\frac{\partial\CW^{D}}{\partial \vec a^{D}} = 0\ ,\label{type B}
}
implying that now $\vec a$ is quantized in units of $\e$.
The two quantization conditions have been called type A and type B, respectively. Due to (\ref{legendre}), the twisted superpotential and its dual are equal as multi-valued functions when either type A or type B quantization condition is imposed. 

An earlier observation of Nekrasov and Shatashvili \cite{Nekrasov:2009uh, Nekrasov:2009ui} also relates the twisted superpotentials $\CW^\text{2d}$ of certain two-dimensional gauged linear sigma models to the Yang-Yang functionals of quantum integrable systems. Here, the given two-dimensional theory acquires a twisted mass $\e$ from the $\Omega$ background, and the F-term equation is then identified with the Bethe ansatz equation of the quantum integrable system. The supersymmetric vacua of the two-dimensional theory are in one-to-one correspondence with the eigenstates of the integrable system. 
These proposals have been powerful not only at establishing connection between individual supersymmetric field theories and quantum integrable systems, but they also provide a novel way to relate gauge theories in various spacetime dimensions by relating them to the same integrable system \cite{Orlando:2010uu, Muneyuki:2011qu}.

Based on these ideas, \cite{Dorey:2011pa, Chen:2011sj} proposed and proved a duality between four-dimensional $\mathcal{N}=2$ superconformal QCD and two-dimensional $\mathcal{N}=(2,2)$ supersymmetric sigma model by relating them to the quantization of the $sl(2, \BR)$ spin chain, where the two-dimensional theory can be viewed as the world-volume
theory of vortex strings probing the root of the Higgs branch of the
four-dimensional theory. The vortices carry quantized units of
magnetic flux and give rise to the type B quantization of the
four-dimensional theory. The precise statement of the duality is that
the twisted superpotentials of the two theories coincide when evaluated on their 
corresponding vacua; loosely speaking
\eq{
\CW^\text{4d} \quad \stackrel{\text{on-shell}}{\cong} \quad \CW^\text{2d}\ .
}

One can naturally lift this duality to one higher dimension and connect theories with fundamental flavors in five and three dimensions. In this paper we explore more general theories with massive adjoint hypermultiplets. The theories with fundamental flavors can be recovered via decoupling limits of the quiver generalization.

{\bf Five-dimensional theory}: $\CN=2^{*}$ $U(L)$ super Yang-Mills theory on $\BR^{4} \times S^{1}$ with an adjoint hypermultiplet of mass $m$.
Its low energy effective theory is four-dimensional with gauge coupling $1/g^{2}_\text{4d} = 2\pi R/g^{2}_\text{5d}$, which is combined with the vacuum angle $\th_\text{4d}$ to form an effective coupling $\t = 4\pi i/g^{2}_\text{4d} + \th_\text{4d}/2\pi$. The theory is subject to the Nekrasov-Shatashvili deformation in $\BR^{4}$. 

{\bf Three-dimensional theory}: $\CN=4$ $U(N)$ super Yang-Mills theory on $\BR^{2} \times S^{1}$ with three adjoint chiral multiplets of twisted masses $m$, $\e$, and $-m-\e$. The 3d Fayet-Iliopoulos parameter $\z$ is related to the 2d parameter $r$ as $\z = r/(2\pi R)$. It is combined with the vacuum angle $\th_\text{2d}$ as a complex coupling $\t = ir + \th_\text{2d}/2\pi$.

In the same spirit as \cite{Dorey:2011pa}, the three-dimensional theory arises as the world-volume theory of vortex strings probing the Higgs branch of the five-dimensional theory. One of the new features of these odd dimensional theories is the possibility of adding Chern-Simons terms. As usual they can be induced by integrating out massive fermions. As we will show, turning on a level $k$ five-dimensional Chern-Simons term then is related by the duality to a three-dimensional Chern-Simons term of the same level. In the ensuing dynamics, the vortices become dyonic carrying electric as well as magnetic fluxes.

The duality is established by relating the two theories to the same integrable system, in this case the elliptic Ruijsenaars-Schneider model \cite{Ruijsenaars:1986vq}. This is a relativistic\footnote{Note that the name ``relativistic'' is somewhat misleading and it would be better to call it ``one-parameter generalization'' \cite{Braden:1997nc}.} generalization of the elliptic Calogero-Moser model describing $L$ particles on a circle of radius $\t$ interacting via a pairwise elliptic potential having a coupling proportional to $m$. It is known that the spectral curve of the model coincides with the Seiberg-Witten curve associated with the five-dimensional theory \cite{Nekrasov:1996cz, Braden:1999zpa}. We find that the saddle-point equation gives a Bethe ansatz equation describing the magnon excitations of this integrable model. 

We also generalize our 5d/3d duality to the elliptic quiver theory and consider its various limits.
When the adjoint or bi-fundamental hypermultiplet masses become very heavy, we can integrate them out and the theory becomes pure five-dimensional $\CN=1$ super Yang-Mills corresponding to the relativistic Toda theory. When the gauge couplings of the two ends of the quiver tend to zero, the corresponding gauge groups freeze and become flavor groups and we obtain five-dimensional $\CN=2$ supersymmetric QCD or its linear quiver generalization which correspond to the anisotropic XXZ integrable spin chain. Finally in the $R \to 0$ limit, we also recover the four-dimensional theories corresponding to the non-relativistic Toda chain and XXX spin chain, respectively. 

This paper is organized as follows. In section 2, we describe the various five-dimensional gauge theories under consideration and their brane constructions. In section 3, we discuss in detail the related three-dimensional gauge theories. In section 4, we provide evidence for the 5d/3d duality through perturbative and non-perturbative checks, and demonstrate how the Bethe ansatz equations naturally appear from the non-perturbative part of the five-dimensional Nekrasov partition function. We conclude by commenting on some future directions.

\section{Five-dimensional Theory}

\subsection{\texorpdfstring{$\CN=2^{*}$}{N=2*} super Yang-Mills theory on \texorpdfstring{$S^{1}$}{S1}}

Let us begin by reviewing the brane construction of a supersymmetric gauge theory with eight real supercharges compactified on ${\mathbb R}^4\times S^1$, 
a theory that can be regarded as the five-dimensional generalization of the familiar four-dimensional $\CN=2^{*}$ theory \cite{Dorey:2004xm}. A perhaps less-known but important fact is that this theory also admits Higgs branches when the mass parameter is tuned to special values and this implies the existence of co-dimension two vortex strings probing the Higgs branch and motivates the 5d/3d correspondence in our subsequent discussions.

We start from the construction of $\CN=(1,1)$ super Yang-Mills theory in five dimensions with gauge group $U(L)$. This theory is maximally supersymmetric with sixteen supercharges and can be realized in IIB string theory with $L$ coincident D5 branes extended in the 012346 directions. Two of the directions are compactified on a two-torus given by $x^4 \sim x^4 +2\pi R$ and $x^6 \sim x^6 + 2\pi R_6$ {}. At energy scale $\Lambda \gg 1/R_6 \gg 1/R$, we have an effective five-dimensional theory on $\BR^4\times S^1$. By T-dualizing on the $x^{4}$ circle, we obtain an equivalent IIA brane construction with the $L$ D4 branes localized at a point on the dual circle of radius $1/R$. We can also separate the D4 branes along the compact $x^4$ direction and this corresponds to turning on a Wilson line for the $x^4$-component of the gauge field in the five-dimensional theory. Upon compactification, the Wilson line yields a periodic scalar that combines with a real adjoint scalar in the five-dimensional vector multiplet to form a complex scalar. Geometrically its VEV gives the positions of individual D4 brane in the complex direction $u=x^5+ix^4$ where $u \sim u+i 2\pi/R $. 
To obtain the five-dimensional theory of interest, we need to break the supersymmetry by half, and this is achieved by introducing a single NS5 brane in the 012345 directions where the D4 branes can break once $x^6$-periodicity is imposed. We also need to give a complex mass to the adjoint hypermultiplet, this can be done following \cite{Witten1997} by giving the D4 brane twisted boundary condition: $x^6 \to x^6 + 2\pi R_6$, $u \to u+ m$, such that the two end points of a D4 brane on the NS5 brane no longer meet but separate by an amount $2\pi R m$ in complex $u$ plane.
Finally the compactification along the $x^4$ direction further demands the mass parameter $m$ to satisfy the condition $m \sim m+2\pi/R$.
This finishes the brief review on brane construction of the five-dimensional theory.
 
\begin{figure}[ht]
\begin{center}
\includegraphics[scale=0.8]{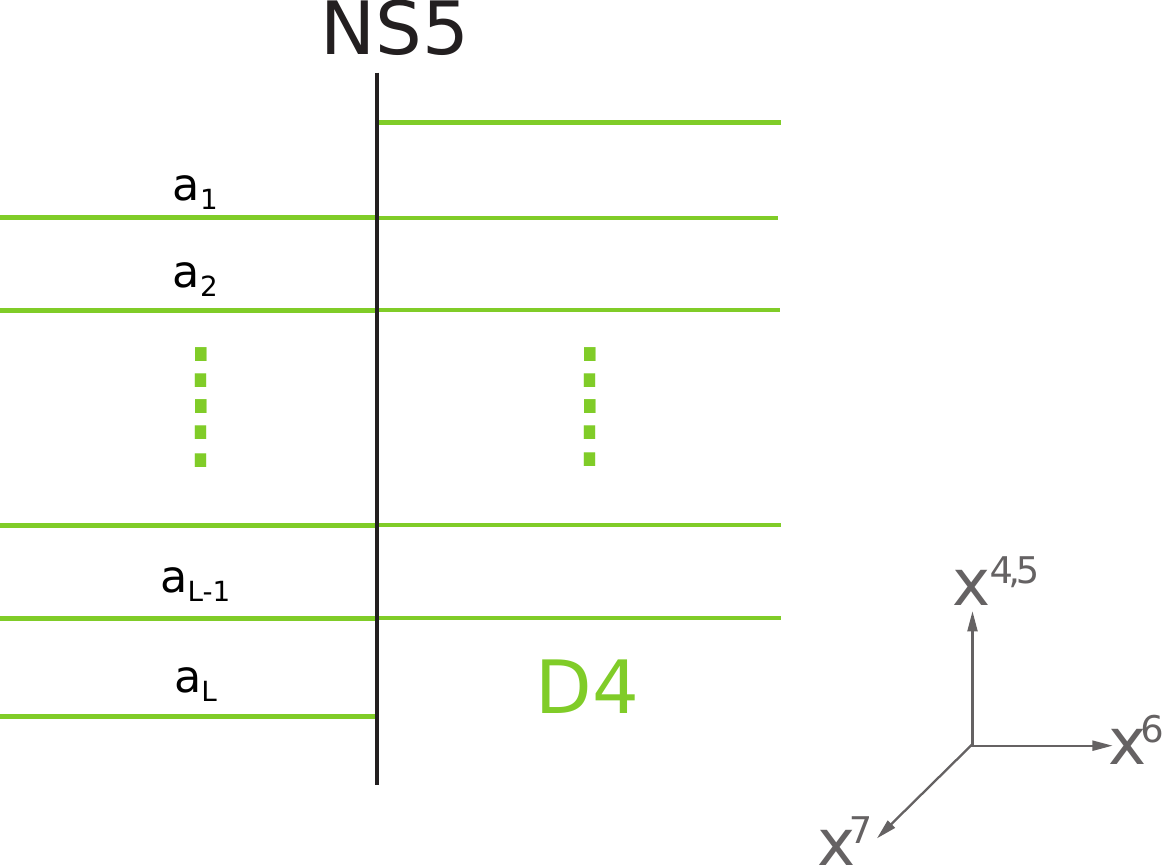}
\caption{The Higgs branch root with adjoint hypermultiplet mass $m$.}
\label{connect}
\end{center}
\end{figure}

Let us denote the coordinates on the Coulomb branch of the five-dimensional theory by $a_i$, which can be regarded as the diagonal VEVs of the adjoint scalar in the vector multiplet. The periodicity in the $x^4$ direction can be realized by imposing $a_i = a_{i+L}$. If the gauge group is broken down to its maximal Cartan subgroup $U(1)^L$, 
the D4 branes can now join together and form a single helical D4 brane or a coil wrapping along the two-torus given by the $x^4$ and $x^6$ directions as illustrated in Figure \ref{connect}. 
We can identify the $i^\text{th}$ circular segment of the helical D4 brane with the $i^\text{th}$ $U(1)$ abelian factor of $U(1)^L$.
We therefore view the five-dimensional world volume of the D4 brane in such configuration as stretching along the non-compact $0123$ directions and wrapping along the $(x^4, x^6)$ torus.

As mentioned before, our five-dimensional theory also admits a Higgs branch which meets the Coulomb branch at special loci known as the ``Higgs branch root'' given by 
\eq{
a_{j} - a_{j+1} = m, \qquad j=1,\dots, L\ ,\label{HiggsRoot}
}
up to permutations of the $\{a_j\}$.
Now if we quantize the mass parameter via 
\eq{
mLR \in 2\pi \mathbb{Z} \label{higgs},
}
then the two ends of the helical D4 brane are identified and we obtain a single toroidal helical D4 brane. Using the $U(1)$ center-of-mass degree of freedom, we may set $a_j=-jm$ and the condition $a_j=a_{j+L}$ is consistent with the periodicity of the mass parameter $m \sim m+2\pi/R$. We can then move onto the Higgs branch of the theory by lifting the single D4 brane in the $x^{7}$ direction away from the NS5 brane.

Following Nekrasov and Shatashvili \cite{Nekrasov2009A}, we now turn on the $\Omega$-deformation in the non-compact $23$ plane. The Higgs branch condition is then deformed into 
\eq{
a_{j} - a_{j+1} = m - (n_{j} - n_{j+1})\e \label{quantization},
}for $n_j\in\mathbb Z$. These are to be viewed as 
a quantum deformation of the classical Higgs branch root condition for the five-dimensional theory.
Here, the integers $n_j$ also have the physical interpretation of turning on quantized magnetic flux in the non-compact $23$-directions for the $j^\text{th}$ $U(1)$ factor in the Cartan subalgebra of $U(L)$:
\eq{\frac{1}{2\pi}\int (F_{23})_{j}~ dx^{2}\wedge dx^{3} = n_{j} \ .\label{defni}} 
 
\begin{figure}[ht]
\begin{center}
\includegraphics[scale=0.8]{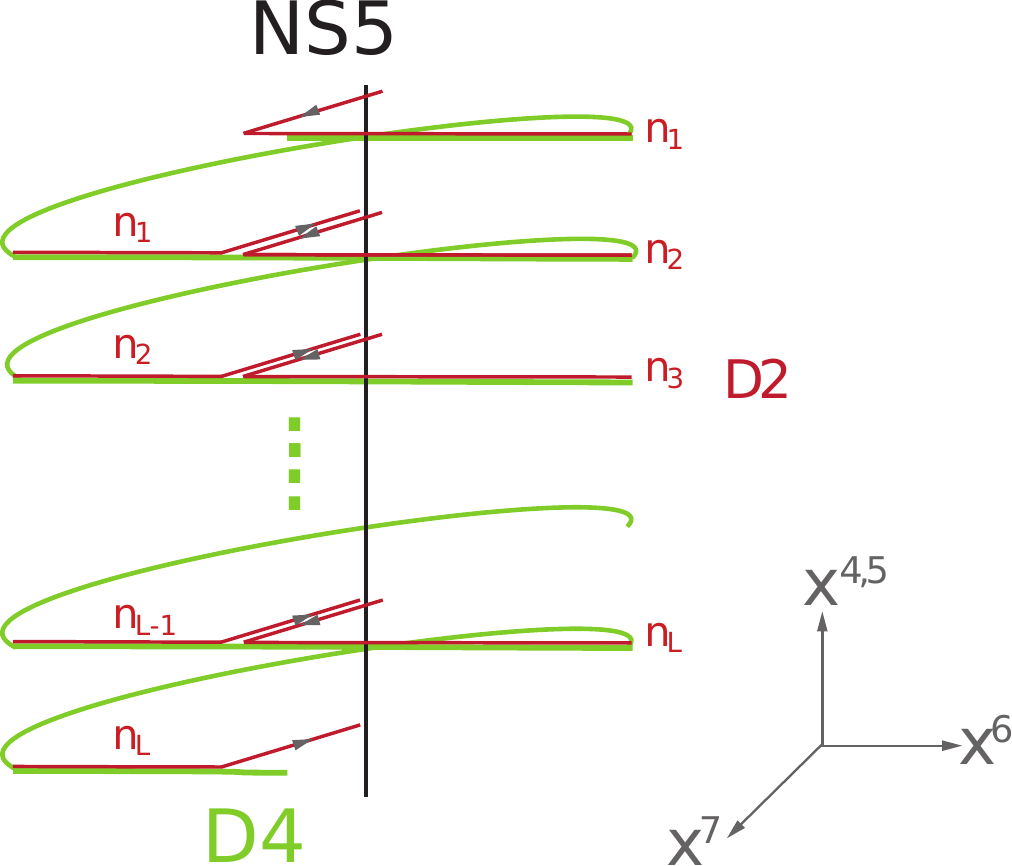}
\end{center}
\caption{Higgs branch of the five-dimensional theory as a single helical D4 brane.} 
\label{helical}
\end{figure}

In the brane construction we can realize such quantized magnetic flux by ``dissolving''
$n_j$
D2 branes in the $j^\text{th}$ D4 brane. However, due to the compact toroidal D4 world volume, the D2 branes are rather subtly engineered.
Essentially, to minimize their energy the D2 branes like to stay inside the resulting single D4 brane. However, due to the opposite orientations of the adjacent D4 branes, this can only be consistent with $n_{j+1}-n_j$ D2 branes stretching from the NS5 brane to the D4 branes. In order to compensate for the net magnetic charges in the D4 brane, as given in (\ref{defni}), we can configure the D2 branes in the following way. Let us begin with $N$ D2 branes and take a subset of  $n_j$ of them and stretch them from the NS5 brane to the D4 brane along the $x^{7}$ direction, then stretch them along the toroidal $x^{4}$ and $x^6$ directions inside the D4 brane world volume at fixed $x^{7}$ and eventually stretch back along $x^7$ to the NS5 brane. This is shown in Figure \ref{helical}. 

Near the region transverse to the helical D4 brane, there are $n_j$ D2 branes of one orientation along $x^7$ and $n_{j+1}$ of the opposite orientations and so they locally annihilate to leave $n_{j+1}- n_j$ D2 branes stretching along the interval.
One important detail here is that as far as the $j^\text{th}$ circular segment of the helical D4 brane is concerned, it only contains $n_j$ unit of net D2 or ``vortex'' charges, this motivates the quantization condition on the Coulomb coordinate:
\eq{
a_j = -jm -\left(n_j+\frac12\right) \epsilon, \qquad i=1,\dots, L\ , \label{quantization2}
}
where we identify $n_j=n_{j+L}$. 
We see that the Higgs branch root condition (\ref{quantization}) is identically satisfied.
Notice that as the quantization condition (\ref{quantization}) only fixes $a_j$ up to an additive constant, we have used this freedom to introduce an additional $\epsilon/2$. This accounts for the difference between the five-dimensional and three-dimensional Coulomb parameters in the deformed background.
In section 3, we shall discuss in some detail the three-dimensional world-volume theory of these D2 branes or ``vortices'' from the perspective of the five-dimensional gauge theory. 

\subsection{Elliptic quiver generalization}
It is straightforward to extend the analysis to the five-dimensional generalization of the elliptic quiver theory considered in \cite{Witten1997}.
The quiver gauge group is then $G=U(1)\times SU(L)^K$ where the $U(1)$ factor corresponds to the center-of-mass motion and decouples from the rest. The matter content of the $\a^\text{th}$ factor $SU(L)_\a$ consists of a bi-fundamental hypermultiplet ${\bf L}_{\a} \otimes \overline{\bf L}_{\a+1}$ with mass $m_\a$.
The Coulomb branch coordinates are $a_{j}^{\a}$ with $\a=1, \dots, K, ~ i=1,\dots, L$ and they have periodicity $a^{\a+K}_{j} \sim a^\a_{j}+m$. The bi-fundamental mass $m_{\a}$ is given by the difference between the center-of-mass positions of the D4 branes ending on either side of the $\a^\text{th}$ NS5 brane; that is
$m_\a=\frac{1}{L}\sum_{j=1}^{L}\left(a_{j}^{\a+1}-a_{j}^{\a}\right)$.
The constraint $\sum_{\a=1}^{K} m_{\a}=0$ is automatically satisfied by $m_\a$ as defined above. However, as explained in \cite{Witten1997}, this can be relaxed to give $\sum_{\a=1}^K m_\a = m$, as in the single gauge group case, by having a non-trivial twisted fibration along the compactified $x^6$ direction in the brane construction. In this way, one can have arbitrary hypermultiplet masses $\{m_\a\}$.

\begin{figure}[ht]
\begin{center}
\includegraphics[scale=0.8]{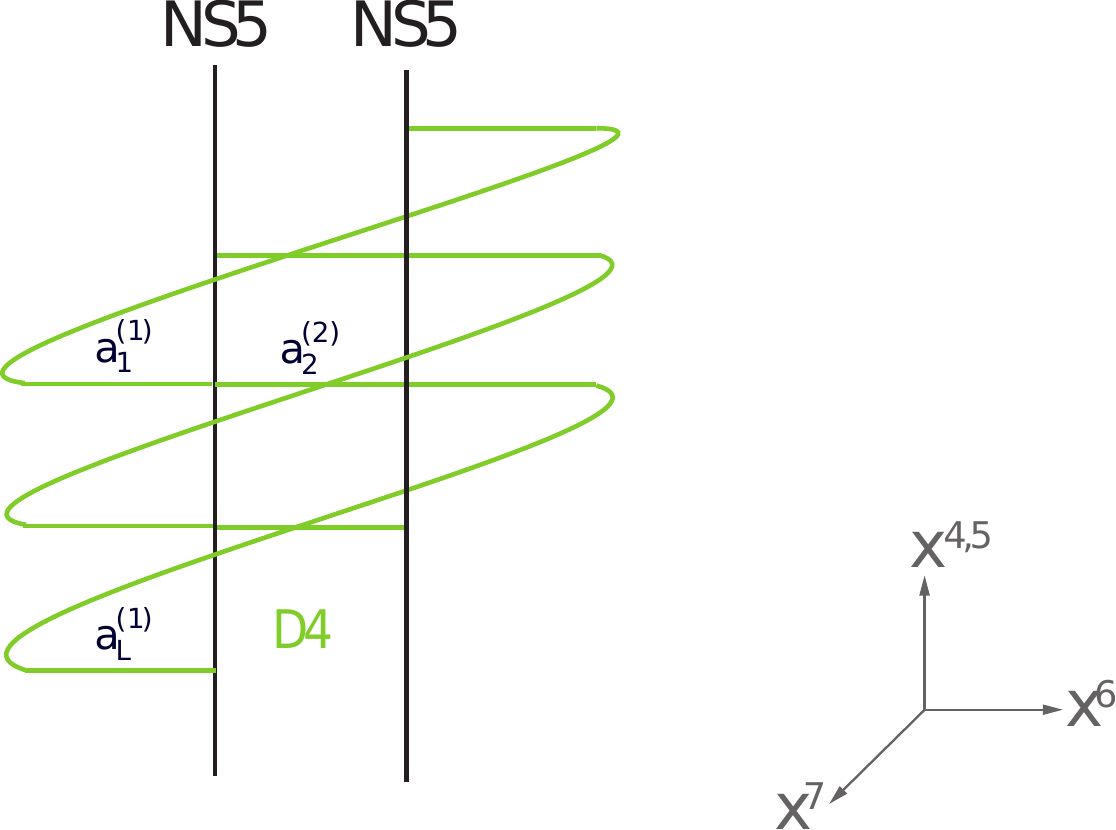}
\caption{The Higgs branch root of the $U(1) \times SU(3)^{2}$ elliptic quiver theory.}
\label{elliptic5d}
\end{center}
\end{figure}
\begin{figure}[ht]
\begin{center}
\includegraphics[scale=0.8]{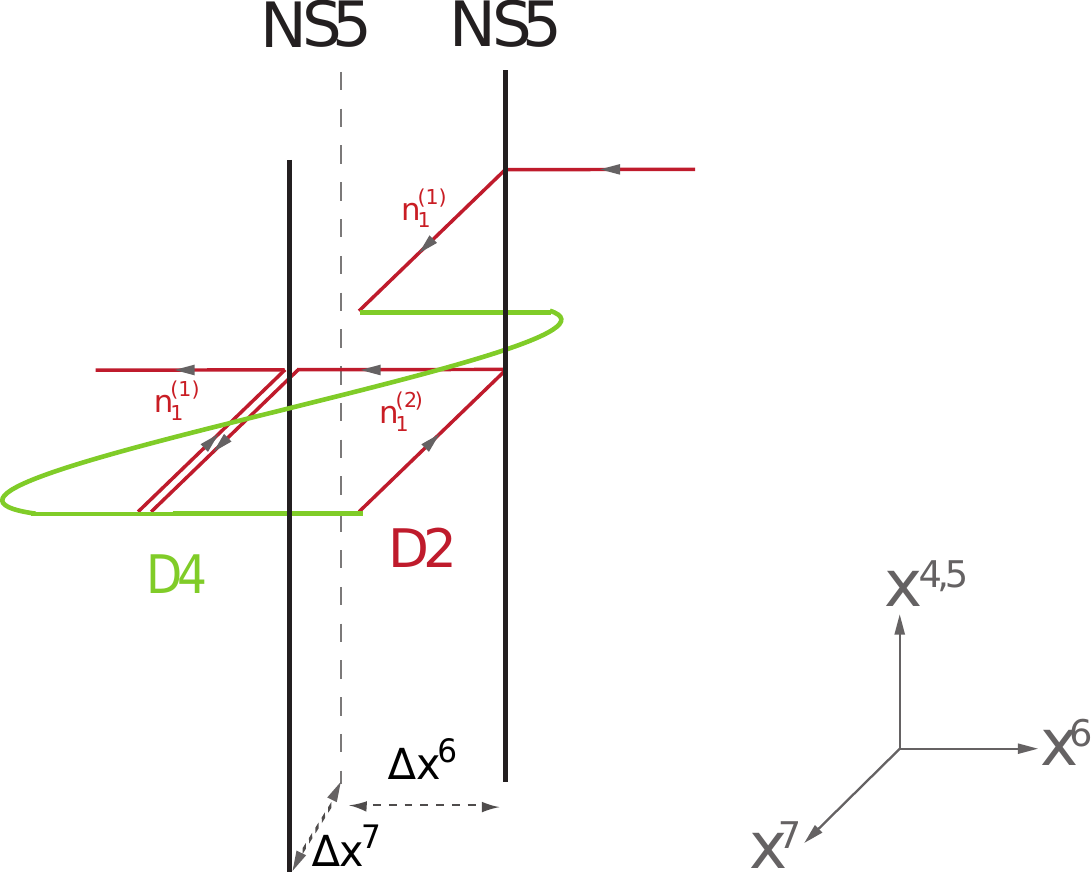}
\caption{Three-dimensional $U(N)^{2}$ theory dual to five-dimensional $U(1) \times U(1)^{2}$ theory.}
\label{elliptic3d}
\end{center}
\end{figure}

The Higgs branch root in the quiver generalization is located at
\eq{a^{\a}_{j}-a^{\a+1}_{j+1} = -m_{\a}.}
As explained in \cite{Dorey:2001qj}, the D4 branes will connect to form $K$ helices each beginning on one NS5 brane and ending on another NS5 brane. If $K$ and $L$ are co-prime, then each helix connects two different NS5 branes. If we now quantize the bi-fundamental masses in the form
\eq{m_{\a}LR \in 2\pi \mathbb{Z}\label{elliptichiggs},}
the helices connect to form a single torus helix and the gauge group is broken to $U(1)^{\gcd(K,L)}$. If $K$ and $L$ are not co-prime, then each helix begins and ends on the same NS5 brane. Thus we get $\gcd(K,L)$ torus helices when we impose (\ref{elliptichiggs}). Figure \ref{elliptic5d} shows an example with $K=2$ and $L=3$.
In either case, we can dissolve $n^{\alpha}_{j}$ D2 branes in the $j^\text{th}$ segment of the helical D4 brane(s) connected to the $\a^\text{th}$ and the $(\a+1)^\text{th}$ NS5 branes and move the D4 brane(s) away from the NS5 branes leaving $n^{\a}_{j}$ D2 branes stretched in $x^{7}$ between the $\a^\text{th}$ NS5 brane and the $j^\text{th}$ segment of the D4 brane(s). Between the $\a^\text{th}$ and the $(\a+1)^\text{th}$ NS5 branes, there are $\sum^{L}_{j=1}n^{\a}_{j} = N$ D2 branes. Figure \ref{elliptic3d} shows a simple example.
Note that the $N$ D2 branes are stretched in $x^{6}$ as well as $x^{7}$. The FI parameter $r$ is proportional to $\Delta x^{6}$ and parametrizes the Higgs branch VEV of the three-dimensional theory. The five-dimensional Higgs branch VEV corresponds to $\Delta x^{7}$. 
When $r > 0$, the three-dimensional theory lies on the Higgs branch.

The quantum deformation away from the Higgs branch root in the quiver is now given by the condition
\eq{
a^{\a}_{j} - a^{\a+1}_{j+1} = -m_{\alpha}-\left(n^{\a}_{j} - n^{\a+1}_{j+1}\right)\e\ .
}

\subsection{Adding Chern-Simons terms}

In addition to the quiver generalization, the five-dimensional gauge theories can be endowed with Chern-Simons terms of the form \cite{Intriligator:1997pq}
\eq{
\mathcal{L}_\text{CS}= \frac{c_{ijk}}{24\pi^{2}} A^{i} \wedge F^{j} \wedge F^{k}. \label{CSterm}
}
The well-known way of introducing a Chern-Simons term is to begin with the five-dimensional gauge theory without one and couple the theory with $k$ auxiliary massive fundamental fermions, each of mass $\m$. The parity triangular anomaly is generated via a one-loop radiative correction involving the auxiliary massive fermions. The net effect is to induce a Chern-Simons term in the low energy effective action with $c_{ijk} \sim -k/2\cdot \text{sgn}(\m)$ when $i=j=k$. 
In the IIA brane picture, the additional auxiliary massive fermions can be realized by having D6 branes in $0123789$ directions and integrating out the open strings stretching between D4 and D6 branes. The scenario can also realized by a brane set-up in the IIB theory \cite{KLL}.

In the presence of the Chern-Simons term, topologically stable vortex solutions exist in the so-called ``broken phase'' that carry both electric and magnetic charges (see \cite{CSreview} for an excellent introduction to these solitons). Their world-volume dynamics can also be studied by replacing the bulk Chern-Simons term with auxiliary fermions \cite{CSvortex1, CSvortex2}. These fermions arise as the zero mode fluctuations of the bulk auxiliary fermions. Given that there is a simple way to introduce a Chern-Simons term we do not expect that its presence affects the quantization condition (\ref{quantization1}). We shall confirm this expectation by performing an explicit check of the 5d/3d duality in the presence of a Chern-Simons term in section 4.
Here we also mention that in three dimensions, the IIB brane configurations for the vortices in supersymmetric Chern-Simons theory with fundamental hypermultiplets have been realized in \cite{KOhta, LLOY}. It would be interesting to consider its generalization to incorporate adjoint hypermultiplets instead and our configuration should then be obtained via a T-duality transformation.

\section{Three-dimensional Theory}

\subsection{\texorpdfstring{$\mathcal{N}=4$}{N=4} super Yang-Mills theory on \texorpdfstring{$S^{1}$}{S1}}

Having discussed the five-dimensional theories of interest, 
let us now identify the theory describing the collective excitations of the set of $\sum_{i=1}^{L} n_{i} = N$ D2 branes or vortices. We shall do so by returning to the IIB picture that has a stack of $N$ D3 branes stretched in the 0146 directions and intersecting a single NS5 brane in the 012345 directions and, for the moment, we turn off the $\Omega$-deformation.
As before, we compactify the configuration on the $(x^{4},x^{6})$ torus with the radii $R\gg R_6$ and study the resultant effective theory on $\BR^{2} \times S^{1}$.\footnote{To recover the IIA picture, one has to  T-dualize along the $x^{4}$ circle and separate the D2 branes by turning on Wilson lines as before.}
The world-volume theory is three-dimensional $\mathcal{N}=4$ super Yang-Mills theory with gauge group $U(N)$, whose matter content includes a vector multiplet $V$, an adjoint hypermultiplet $H$ and $L$ fundamental hypermultiplets $H_{i}$, all massless. Because the three-dimensional theory is compactified on $S^{1}$, the effective two-dimensional theory preserves $\mathcal{N}=(4,4)$ supersymmetry. In terms of $\mathcal{N}=(2,2)$ language, the three-dimensional $U(N)$ vector multiplet $V$ decomposes into an adjoint twisted chiral multiplet $\Sigma = \overline D_{+}D_{-}V$ and an adjoint chiral multiplet $\Phi$. The real scalar in $V$ combine with the Wilson line of $A_{4} \in V$ around the compact circle to form the complex scalar $\l \in \Sigma$. The three-dimensional adjoint hypermultiplet $H$ decomposes into a pair of adjoint chiral multiplets $K, \widetilde K$ and the $L$ fundamental hypermultiplets into $L$ fundamental chiral multiplets and $L$ anti-fundamental chiral multiplets. Each field is accompanied by an infinite tower of massive Kaluza-Klein modes.

Now we turn on the Nekrasov-Shatashvili version of the $\Omega$-deformation in the $23$ plane. The resulting theory has $\CN=(2,2)$ supersymmetry in the undeformed $(x^0,x^1)$ plane and there are twisted mass for the fields transforming under the $(x^2,x^3)$ rotation. The adjoint chiral fields $\Phi, K, \widetilde K$ have rotation charge $1, -\frac12, -\frac12$, respectively. Such rotational symmetry can also be mixed with the global $U(1)_R$-symmetry, under which the fields are charged $0, \frac12, -\frac12$, respectively. Under the combined rotations, $\Phi$, $K$ and $\widetilde{K}$ acquire twisted masses $\e, 0$ and $ -\e$, respectively. The brane construction requires us to impose the twisted periodic condition $x^6 \to x^6 + 2\pi R_6$, $u \to u+ m$, which gives additional twisted masses $0, m$ and $ -m$ to $\Phi, K$ and $\widetilde{K}$. We can also introduce the three-dimensional Fayet-Iliopoulos parameter $\z$, which is related to the two-dimensional Fayet-Iliopoulos parameter $r$ by $r = 2\pi R \z$. It combines with the two-dimensional vacuum angle $\th_{\rm 2d}$ to give rise to a complex coupling $\t = ir + \th_\text{2d}/{2\pi}$ which will be identified with the five-dimensional coupling.
To summarize, we obtain an effective two-dimensional $\mathcal{N}=(2,2)$ theory with $U(N)$ gauge group and coupling $\t$. Its field content consists of three adjoint chiral multiplets with masses $\e,m, -m-\e$, $L$ massless fundamental and $L$ massless anti-fundamental chiral multiplets along with a tower of massive Kaluza-Klein modes for each field. 

The low energy theory is described by an effective twisted superpotential $\CW^\text{3d}$, whose tree level contribution includes the Fayet-Iliopoulos parameter and the theta angle \cite{Aganagic:2001uw}
\eq{\CW^\text{3d}_\text{FI} = 2\pi i \t\Sigma\ .}
In addition, we can introduce a level $k$ Chern-Simons term, which contributes to the tree level action the term 
\eq{\CW^\text{3d}_\text{CS} = - \frac{Rk}{2} \Sigma^{2}.}
The one-loop contribution from each adjoint chiral field with twisted mass $m$ includes sum over the infinite tower of massive Kaluza-Klein modes
\eq{\CW^\text{3d}_\text{adj} = \sum_{n\in\mathbb{Z}} \left(\Sigma - m + \frac{2\pi n}{R}\right) \left[\log \left(\Sigma - m + \frac{2\pi n}{R}\right) - 1\right]\ .}

In summary, when we integrate out the massive fields, the effective twisted superpotential is written as a sum over the scalars $\l$ in the chiral multiplet $\Sigma$.
\EQ{
\CW^\text{3d}&= 2\pi i \t \sum_{a=1}^{N} \l_{a}-\frac{Rk}{2} \sum_{a=1}^{N} \l^{2}_{a}+ 
\sum_{a,b=1}^{N}\sum_{n\in \IZ}\left[f\left(\l_{ab} - \e - \frac{2\pi n}{R}\right) \right.\\
&\left. + f\left(\l_{ab} - m - \frac{2\pi n}{R}\right) + f\left(\l_{ab} + m+\e - \frac{2\pi n}{R}\right)\right],
\label{superpotential}}
where $f(x) = x(\log x - 1)$ and we define $\l_{ab}=\l_{a} - \l_{b}$. 
The vacuum of the theory is described by the set of F-term equations
\eq{\exp\Big[\frac{\partial\CW^\text{3d}}{\partial \l_{a}}\Big] = 1\ .}
It coincides with a set of Bethe ansatz equations which determines the eigenstates of a quantum integrable system. The eigenstates are labelled by a set of magnon rapidities, which we identify with the on-shell value of the complex scalars $\l_{a}$.
For the theory under consideration, its vacuum equations are
\eq{
e^{Rk\l_{a}}=-e^{2\pi i \t}\prod_{b=1}^{N}\frac{\sin\frac{R}{2}\left(\l_{ab}-\e\right)\sin\frac{R}{2}\left(\l_{ab}-m\right)\sin\frac{R}{2}\left(\l_{ab}+m+\e\right)}{\sin\frac{R}{2}\left(\l_{ab}+\e\right)\sin\frac{R}{2}\left(\l_{ab}+m\right)\sin\frac{R}{2}\left(\l_{ab}-m-\e\right)}.
\label{bae}}
In the mapping to the quantum integrable system, we identify the parameter $\e$ with Planck's constant via $\e = -i\hbar$.  The magnons undergo factorizable scattering 
characteristic of an integrable system
whose total scattering phase is the product of two-body S-matrices as given in the right-hand side of (\ref{bae}). The additional phase $e^{2\pi i \t}$ represents a twisted periodic boundary condition of the scattering. The factor from the Chern-Simons term is somewhat harder to interpret in the integrable model and will be explained in detail later. The total number of magnons coincides with the number of vortices. Note that unlike the purely two-dimensional theory whose F-term equation is a set of rational functions, the equation is trigonometric due to the sum over an infinite tower of Kaluza-Klein modes. Note that the hypermultiplets do not contribute to the F-term equation. 

We can recover the two-dimensional limit by taking $R \to 0$. Note that the twisted mass $m$ must be taken to infinity because the mass parameter is tuned as (\ref{higgs}). The Bethe ansatz equation assumes the more familiar rational form
\eq{
1=-q\prod_{b=1}^{N}\frac{\l_{ab}-\e}{\l_{ab}+\e}\ .
\label{bae4d}}
It is precisely the F-term equation for the four-dimensional $\mathcal{N}=2^{*}$ $SU(N)$ theory with an adjoint chiral field of mass $\e$.
By examining the equation, we conclude that there is at most one magnon solution above the vacuum. To see this, let us rewrite (\ref{bae4d}) as the following polynomial equation
\eq{\prod_{b=1}^{N}\left(\l - \l_{b}+\e\right)+q\prod_{b=1}^{N}\left(\l - \l_{b}-\e\right)=\left(1+q\right)\prod_{b=1}^{N}\left(\l - \l_{b}\right)\ .}
Defining $Q(\l) = \prod_{a=1}^{N}(\l-\l_{a})$, the above can be written as a periodicity condition for the polynomial $P(\l) = Q(\l+\e)-Q(\l)$,
\eq{P(\l) = qP(\l-\e)\ . \label{periodic}}
The only polynomial solution to (\ref{periodic}) is a constant, which is nonzero only when $q=1$.
This implies that $Q(\l)$ also satisfies a periodicity condition for some constant $c\e$,
\eq{Q(\l) = Q(\l-\e) + c\e.}
The most general solution is a linear function of $\l$: $Q(\l) = c(\l-\l_{1})$, where the magnon rapidity $\l_{1}$ is an unconstrained complex number.\footnote{We thank Adam Rej for discussion on the argument outlined here.} Recall that the number of magnons coincides with the number of vortices. We conclude that there is no vortex unless $q=1$, i.e.~when the Fayet-Iliopoulos parameter $r$ vanishes. But there is a contradiction here because the Fayet-Iliopoulos parameter must be nonzero for BPS vortices to exist in the first place. The only consistent interpretation is that there
are no BPS vortices in the four-dimensional $\mathcal{N}=2^{*}$ theory.

\subsection{Elliptic quiver generalization}

To generalize to the elliptic quiver case, it is convenient to use a method of images and consider an infinite array of periodic images of the branes in the $x^6$ direction. This gives rise to an infinite quiver gauge theory, but we can recover the original picture by imposing a suitable periodicity constraint on the VEVs of the infinite theory.

Following the logic of the previous section, 
the world-volume theory on the D2 branes in the infinite quiver generalization is three-dimensional $\CN = 4$ super Yang-Mills theory with an infinite product gauge group
$G = \bigotimes^{\infty}_{\a=-\infty} U(N)_{\a}$ compactified on ${\mathbb R}^2\times S^1$. Its matter content consists of an adjoint vector multiplet $V_{\a}$ and a bi-fundamental hypermultiplet $H_{\a,\a+1}$ for each $U(N)_{\a}$, and $L$ fundamental hypermultiplets $H_{i}$. The bi-fundamental hypermultiplet $H_{\a,\a+1}$ carries mass $m_{\a}$, which coincides with the mass of the five-dimensional bi-fundamental hypermultiplet. This is clear from the brane picture because the D2 branes are embedded in the D4 branes and their position coincide. We can rewrite the fields in the $\mathcal{N}=(2,2)$ language as before. The bi-fundamental hypermultiplet $H_{\a, \a+1}$ decomposes into a bi-fundamental chiral multiplet $K_{\a}$ in the representation $\bf N_{\a}\otimes \overline N_{\a+1}$ with mass $m_{\a}$ and another $\widetilde K_{\a}$ in the representation $\bf \overline N_{\a} \otimes N_{\a+1}$ with mass $-m_{\a}$.

Compactifying on the $x^{4}$ circle and turning on the Nekrasov-Shatashvili deformation as before, we find that the $\CN = (2,2)$ adjoint and bi-fundamental chiral multiplets $\Phi_{\a}, K_{\a,\a+1}, \widetilde K_{\a,\a+1}$ carry twisted mass $\e, m_{\a}-\e/2, -m_{\a}-\e/2$, respectively. Each gauge group $U(N)_{\a}$ has its own coupling $q_{\alpha} = e^{2\pi i\t_{\a}}$ and can be endowed with its own Chern-Simons level $k_{\a}$.

The twisted superpotential is simply a sum over each quiver
$\CW^\text{3d} = \sum_{\a=1}^{K}\CW^{\a}$, which receive contributions from the tree level, adjoint and bi-fundamental parts
\EQ{
\CW^{\a}&= 2\pi i\t_{\a} \sum_{a=1}^{N} \l^{\a}_{a}-\frac{Rk_{\a}}{2}\sum_{a=1}^{N} \left(\l^{\a}_{a}\right)^{2}+\sum_{a,b=1}^{N}\sum_{n\in \IZ}\left[f\left(\l^{\a}_{a} - \l^{\a}_{b} - \e - \frac{2\pi n}{R}\right) \right.\\
&\left. + f\left(\l^{\a}_{a} - \l^{\a+1}_{b} - m_{\a} + \frac\e2 - \frac{2\pi n}{R}\right) + f\left(\l^{\a+1}_{a}-\l^{\a}_{b} + m_{\a} + \frac\e2 - \frac{2\pi n}{R}\right) \right].\label{bifundamental}
}
Denoting $Q_{\a}(\l) = \prod_{a=1}^{N}\sin\frac{R}{2}\left(\l-\l^{\a}_{a}\right)$, the F-term equation coincides with the nested Bethe ansatz equation 
\eq{e^{Rk_{\a}\l^{\a}_{a}}=-q_{\a}\frac{Q_{\a-1}\left(\l^{\a}_{a}+m_{\a-1}+\frac\e2\right)Q_{\a}\left(\l^{\a}_{a}-\e\right)Q_{\a+1}\left(\l^{\a}_{a}-m_{\a}+\frac\e2\right)}{Q_{\a-1}\left(\l^{\a}_{a}+m_{\a-1}-\frac\e2\right)Q_{\a}\left(\l^{\a}_{a}+\e\right)Q_{\a+1}\left(\l^{\a}_{a}-m_{\a}-\frac\e2\right)}\ .\label{nBAE}}
To obtain the equation for $K$ quivers, we now impose periodicity in $x^{6}$ by identifying the parameters 
in the $\a^\text{th}$ and the $(\a + K)^\text{th}$ gauge groups as
\EQ{&\l^{\a+K}_{a} = \l^{\a}_{a} + m\ , \qquad m_{\a+K} = m_{\a}\ , \qquad k_{\a+K} = k_{\a}\ , \qquad \t_{\a+K} = \t_{\a}\ .
} 
For $K = 1$, the nested Bethe ansatz equations completely decouple from one another and we recover the Bethe ansatz equation for a single gauge group (\ref{bae}) after identifying $m_{1} = m +\e/2$. The shift in $\e/2$ arises from the additional contribution to the twisted mass from the global $R$-charge that was mentioned earlier.
 
We can deduce the flow to theories with fundamental flavors by considering the decoupling limit $q_{1}, q_{K+1} \to 0$. In this limit, the gauge groups $U(N)_{1}$ and $U(N)_{K+1}$ freeze and the complex scalars are to be treated as constants:
\eq{
\l^{(1)}_{a} = \th_{a}+m_{1}, \qquad \l^{(K+1)}_{a} = \tilde\th_{a}-m_{K}.
}
For the simplest $K=2$ case, the gauge groups $U(N)_{1}$ and $U(N)_{3}$ become flavor symmetries for the $U(N)_{2}$ gauge group in the middle. We obtain three-dimensional super QCD with $N_{f} = 2N$. The integrable model corresponding to this theory is known to be the anisotropic XXZ spin chain. In fact, by moving the terms in (\ref{nBAE}) corresponding to the fundamental flavor contribution to the left-hand side, we obtain
\EQ{
e^{Rk\l_{a}}\prod_{b=1}^{N}\frac{\sin\frac{R}{2}(\l_{a}-\th_{b}-\frac\e2)\sin\frac{R}{2}(\l_{a}-\tilde\th_{b}-\frac\e2)}{\sin\frac{R}{2}(\l_{a}-\th_{b}+\frac\e2)\sin\frac{R}{2}(\l_{a}-\tilde\th_{b}+\frac\e2)}&=q\prod^{N}_{b\ne a}\frac{\sin\frac{R}{2}(\l_{a}-\l_{b}-\e)}{\sin\frac{R}{2}(\l_{a}-\l_{b}+\e) \label{xxzcs}}\ .
}
Without the Chern-Simons term, it is simply the Bethe ansatz equation for the $su(2)$ XXZ spin-$\frac12$ chain of length $2N$ with inhomogeneities $\th_{a}$ and $\tilde \th_{a}$, $a=1, \ldots, N$. Let us interpret the Chern-Simons term from the spin chain perspective. In general, a level $k$ Chern-Simons term can be thought of as the infinite spin limit of a length $k$ XXZ-type spin chain \cite{Gerasimov:2007ap}
\eq{\prod_{\ell=1}^{k} \frac{\sinh\frac{R}{2}(\l_{a} - \th_\ell + s_\ell\e)}{\sinh\frac{R}{2}(\l_{a} - \th_\ell - s_\ell\e)} =q\prod_{b\ne a}^{N}\frac{\sin\frac{R}{2}(\l_{ab}-\e)\sin\frac{R}{2}(\l_{ab}-m)\sin\frac{R}{2}(\l_{ab}+m+\e)}{\sin\frac{R}{2}(\l_{ab}+\e)\sin\frac{R}{2}(\l_{ab}+m)\sin\frac{R}{2}(\l_{ab}-m-\e)}.\label{xxz}}
Once we identify the combination of inhomogeneities and spin with the mass parameters as \cite{Nekrasov:2009uh}
\eq{\m_\ell = -\th_{\ell} + s_\ell\epsilon, \qquad \widetilde \m_\ell = \th_\ell + s_\ell\epsilon,} the left-hand side of (\ref{xxz}) coincides with the F-term equation of $k$ fundamental chiral multiplets with twisted masses $\m_\ell$ and $k$ anti-fundamental chiral multiplets with twisted masses $\widetilde \m_\ell$:
\eq{
\prod_{\ell=1}^{k} \frac{\sinh\frac{R}{2}(\l_{a} - \m_\ell)}{\sinh\frac{R}{2}(\l_{a} - \widetilde \m_\ell)} \to -e^{R(k\l_{a}-\sum^{k}_{\ell=1}\th_\ell)}\ ,
}
as we take $s_\ell \to i\infty$, that is $\m_\ell, \widetilde \m_\ell \to \infty$. This is consistent with the idea that a level $k$ Chern-Simons term can be introduced by coupling $2k$ massive fermions to pure Yang-Mills theory and integrating them out. Thus, to obtain the Bethe ansatz equation (\ref{xxzcs}) including the Chern-Simons term, we can begin with a length $2N+k$ chain and send $k$ of the spins to infinity.

In the two-dimensional $R \to 0$ limit, we obtain the Bethe ansatz equation for the inhomogeneous XXX spin chain.
\EQ{
\prod_{b=1}^{N}\frac{(\l_{a}-\th_{b}-\frac\e2)(\l_{a}-\tilde\th_{b}-\frac\e2)}{(\l_{a}-\th_{b}+\frac\e2)(\l_{a}-\tilde\th_{b}+\frac\e2)}&=q\prod^{N}_{b\ne a}\frac{(\l_{a}-\l_{b}-\e)}{(\l_{a}-\l_{b}+\e)}\ .
}
For general $K$, the Bethe ansatz equation is that of the $su(K)$ XXZ spin-$\frac12$ chain. In the $R \to 0$ limit, we recover the nested Bethe ansatz equation for the $su(K)$ XXX spin-$\frac12$ chain. 

\section{The Duality via Integrability}

In this section, we make the connection between the two theories precise by showing that the twisted superpotentials agree on-shell at their respective vacua. The vacua of the five-dimensional theory are determined by the type B quantization condition (\ref{type B})
\eq{\frac{\partial \CW^{D}}{\partial \vec a^{D}} = 0,}
which implies that the Coulomb branch moduli $\vec a$ are quantized in units of $\e$.
For a single gauge group, they are located at (\ref{quantization2})
\eq{a_{j} = - jm-\left(n_{j}+\frac12\right)\e. \label{quantization1}}
Instead of computing $\CW^{D}$ directly, we use the fact that $\CW = \CW^{D}$ on-shell  to evaluate the twisted superpotential $\CW$ instead. To this end, let us define
\eq{\CG\equiv\left.\CW^{\text{5d}}\right|_{a_{j} = -jm-\left(n_{j}+\frac12\right)\e} - \left.\CW^{\text{5d}}\right|_{a_{j}=-j m - \frac\e2}.}
The vacuum-independent subtraction ensures that the twisted superpotential vanishes at the Higgs branch root when $n_{j} = 0$. The idea is to match this with the three-dimensional twisted superpotential, whose vacua are determined by the F-term equation
\eq{\exp\Big[\frac{\partial \CW^\text{3d}}{\partial \l_{a}}\Big] = 1.}
The vacua are also labelled by $\{n_j\}$, which as we shall see, determine the configuration of the magnons in the complex plane. 
Our goal is to show that
\eq{\CG \thicksim \CW^{\text{3d}}\left(\{n_j\}\right),}
where $\sim$ means equal up to constants that do not depend on the choice of vacuum $\vec n$.
In the brane picture, $n_{j}$ corresponds to the number of D2 branes which are identified with the distance $n_{j}\e$ that the D4 branes are away from the root of the Higgs branch.

The strategy is as follows. Since the Nekrasov partition function can be written as a product of the perturbative part and the instanton part, the twisted superpotential is a sum 
\eq{
\CW^\text{5d} = \CW^\text{5d}_\text{pert} + \CW^\text{5d}_\text{inst}\ .
}
We first check the agreement between the on-shell value of the perturbative part of the twisted superpotentials. This is done by evaluating $\CW^\text{5d}_\text{pert}$ at the quantized value of the Coulomb branch moduli (\ref{quantization1}) and checking that it matches with $\CW^\text{3d}$ evaluated at the leading order solution $\l^{(0)}_{a}$ to the F-term equation, i.e.~the leading order magnon solutions to the Bethe ansatz equation.
\eq{\CG_\text{pert} \sim \CW^\text{3d}(\l^{(0)}_{a})\ .\label{pertcheck}}
Then we show from the instanton partition function that the saddle point equation reproduces the Bethe ansatz equation involving an infinite number of magnons. The quantization condition is equivalent to the truncation to a finite number of magnons. This enables us to obtain directly the Yang-Yang functional from the instanton partition function via
\eq{\CG_\text{inst} = \CW^\text{3d}(\l_{a}) - \CW^\text{3d}(\l^{(0)}_{a})\ , \label{instcheck}}
after imposing the quantization condition.
The check of the proposed duality then follows from (\ref{pertcheck}) and (\ref{instcheck}).

\subsection{Perturbative check}
The perturbative part of the five-dimensional partition function only consists of a classical part and a one-loop part. Without a Chern-Simons term, the classical part is simply
\eq{
\CW^\text{5d}_\text{cl} = -\frac{\log q}{2\e}\sum_{j=1}^{L}a^{2}_{j}\ .}
The Chern-Simons term contributes an additional cubic term \cite{Seiberg:1996bd}
\eq{\CW^\text{5d}_\text{CS} = \frac{Rk}{6\e} \sum_{j=1}^{L} a^{3}_{j}\ ,}
while the one-loop part includes contribution from the tower of Kaluza-Klein modes
\eq{
\CW^\text{5d}_\text{1-loop} = \sum_{i, j=1}^{L}\sum_{n\in \IZ}\left[\w_{\e}\left(a_{i j} + \frac{2\pi n}{R}\right) - \w_{\e}\left(a_{i j} + m + \frac{2\pi n}{R}\right)\right],\label{pert}}
where $\omega_{\e}(x)$ is related to the logarithm of Barnes's double gamma function $\g_{\e_{1},\e_{2}}(x)$ by $\omega_{\e}(x) = \lim_{\e_{2}\to 0}\e_{2}\g_{\e,\e_{2}}(x)$ and this satisfies the identity $\omega'_{\e}(x) = -\log \Gamma(1+x/\e)$.

Evaluating the classical part, without the Chern-Simons term, on shell give
\eq{\CG_\text{cl} \sim -\log q \sum_{j=1}^{L} \left(jm n_{j} + \frac{\e}{2}n_{j}^{2}\right).\label{classical}}
The Chern-Simons term contribution is then
\eq{\CG_\text{CS}\sim -\frac{Rk}{2} \sum_{j=1}^{L} \left[\left(j^2 m^2+j m \e \right) n_j+ \left(j m\e+\frac{\e^{2}}{2} \right) n_j^2+\frac{\e^2}{3}  n_j^3\right]\ .\label{cs}}
The one-loop part can be evaluated using the identity
\eq{\omega_{\e}(x+n\e) - \omega_{\e}(x) = -\e \sum_{s=1}^{n} f\left(\frac{x}{\e}+s\right) - \frac{n\e}{2} \log(2\pi)\qquad \text{for } n>0,}
where $f(x) = x(\log x - 1)$. We will denote $\l^{(0)}_{js} = -jm - s\e$, which will turn out to coincide with the leading-order configuration of magnon roots from the three-dimensional calculation (\ref{lattice}). We find
\eq{\CG_\text{1-loop} = \e \sum_{i,j=1}^{L}\sum_{n\in \IZ} \left(\sum_{s=0}^{n_{ij}-1}-\sum_{s=n_{ij}}^{-1}\right) \left[ f\left(\frac{\l^{(0)}_{(i-j)s}}\e+\frac{2\pi n}{R \e} \right) - f\left(\frac{\l^{(0)}_{(i-j-1)s}}\e+\frac{2\pi n}{R \e}\right)\right],}
where implicitly the contribution from 
a sum starting from a non-negative integer to a negative integer is defined to be zero.
To sum over the Kaluza-Klein modes, we use the identity 
\eq{\sum_{n\in \IZ}f\left(x - \frac{2\pi n}{R}\right) = \frac{i}{R} \left(-\frac{R^{2}}{4} x^{2} + \text{Li}_{2}(e^{iRx})-\frac{\pi^{2}}{6}\right),\label{identity}}
which can be proven by a suitable regulation of the infinite sum and then a limit \cite{Aganagic:2001uw}. The quadratic terms contribute a vacuum-independent constant and can be dropped. We obtain
\eq{\CG_\text{1-loop}\thicksim\frac{i}{R}\sum^L_{i,j=1}\left(\sum_{s=0}^{n_{ij}-1}- \sum_{s=n_{ij}}^{-1}\right)\bigg[\text{Li}_{2}\left(e^{iR\l^{(0)}_{(i-j-1)s}}\right)-\text{Li}_{2}\left(e^{iR\l^{(0)}_{(i-j)s}}\right)\bigg].
\label{1loop}}

Generalizing this result to the elliptic quiver case is straightforward. Using the $U(1)$ center-of-mass degree of freedom, we can choose 
\eq{a^\a_{j} = \sum_{\b=1}^{\a-1}m_{\b}-(j-\a+1)m-\left(n^{\a}_{j}+\frac12\right)\e.} The perturbative part is a sum over contributions from each gauge group $\CW^\text{5d}_\text{pert} = \sum_{\a=1}^{K}\CW^{\a}_\text{pert}$, where
\EQ{
\CW^{\a}_\text{pert} &= -\frac{\log q_{\a}}{2\e}\sum_{j=1}^{L}\left(a^\a_{j}\right)^{2} + \frac{Rk_{\a}}{6\e} \sum_{j=1}^{L}\left(a^\a_{j}\right)^{3} 
\\&- \sum_{i, j=1}^{L}\sum_{n\in \IZ}\left[\w_{\e}\left(a^\a_{i} - a^{\a+1}_{j} + m_{\a}+\frac{2\pi n}{R}\right) - \w_{\e}\left(a^\a_{ij} + \frac{2\pi n}{R}\right)\right]\ .}
The on-shell value of the classical part without the Chern-Simons term is the same as the single gauge group case:
\eq{\CG^{\a}_\text{cl} 
\thicksim -\log q_{\a}\sum_{j=1}^{L}\left[jmn^{\a}_{j} + \frac{\e}{2}(n^{\a}_{j})^{2}\right]\ ,\label{classicalelliptic}}
while the Chern-Simons contribution is
\EQ{
\CG^{\a}_\text{CS} \thicksim
&-\frac{Rk_{\a}}{2}\sum_{j=1}^{L}\left[\left(\sum_{\b=1}^{\a-1} m_{\b}-(j-\a+1)m-\frac\e2\right)^2 n_j^{\a} \right.\\
&\left.-\left(\sum_{\b=1}^{\a-1} m_{\b}\e-(j-\a+1)m\e-\frac{\e^{2}}{2}\right)\left(n_j^{\a}\right)^{2}+\frac{\e^{2}}{3} \left(n_j^\a\right)^{3}\right].\label{cselliptic}
}
The one-loop part can also be calculated in the same way as the single gauge group case above and we obtain
\EQ{
\CG^{\a}_\text{1-loop}&\thicksim
\frac{i}{R}\sum^L_{i,j=1}
\left[
\left( \sum_{s=0}^{n^{\a}_{i}-n^{\a+1}_{j}-1}-\sum_{s=n^{\a}_{i}-n^{\a+1}_{j}}^{-1}\right)\text{Li}_{2}\left(e^{iR\l^{\a(0)}_{(i-j-1)s}}\right)\right.\\
&\left.-\left(\sum_{s=0}^{n^{\a}_{ij}-1}- \sum_{s=n^{\a}_{ij}}^{-1}\right)\text{Li}_{2}\left(e^{iR\l^{\a(0)}_{(i-j)s}}\right)\right],
\label{elliptic1loop}}
where we have used the shorthand notation 
\eq{\l^{\a(0)}_{js} = \sum_{\b=1}^{\a-1}m_{\b}-(j-\a+1)m- s\e\ ,}
quantities which will turn out to coincide with the leading order solution to the nested Bethe ansatz equation (\ref{lattice1}).

Let us now move to the three-dimensional theory. Recall the twisted superpotential for a single gauge group (\ref{superpotential}), which we now sum over the Kaluza-Klein modes using (\ref{identity}) 
\EQ{
\CW^\text{3d} &\thicksim 2\pi i\t \sum_{a=1}^{N} \l_{a}-\frac{Rk}{2} \sum_{a=1}^{N} \l^{2}_{a} \\& -\frac{i}{R}\sum_{a,b=1}^{N}\left\{\frac{R^{2}}{4}\Big[\left(\l_{ab}-\e\right)^{2} + \left(\l_{ab}-m\right)^{2} + \left(\l_{ab}+\e+m\right)^{2}\Big]\right.\\
&-\text{Li}_2\left(e^{i R(\l_{ab}-\e)}\right) -\text{Li}_2\left(e^{i R(\l_{ab}-m)}\right)-\text{Li}_2\left(e^{i R(\l_{ab}+\e+m)}\right)\!\bigg\}\ .\label{3dsuperpotential}
} 
To evaluate the twisted superpotential (\ref{superpotential}) on-shell, we need to solve the vacuum equation (\ref{bae}) order-by-order in $q$ expansion. Let us write the solution as $\l_{a} = \l^{(0)}_{a} + q\l^{(1)}_{a} + \cdots$. To the leading order, we have to solve
\eq{\prod_{b=1}^{N}\sin \frac{R}{2}\left(\l^{(0)}_{ab}+\e\right)\sin\frac{R}{2}\left(\l^{(0)}_{ab}+m\right)\sin\frac{R}{2}\left(\l^{(0)}_{ab}-m-\e\right) = 0.}
The magnons form a lattice in the complex plane separated by $m$ along the real axis and $\e = -i\hbar$ along the imaginary axis. 
We label them by
\eq{\l^{(0)}_{jq} = -jm - q\e \mod \frac{2\pi}{R} \label{lattice},}
where $j=1,\ldots, L$, and  $q=1,\ldots, n_{j}$ with $n_{i} > n_{j}$ for $i > j$.
Evaluating on-shell, the tree level part agrees with the classical part in the five-dimensional theory (\ref{classical}, \ref{cs}).

The on-shell value of the part coming from the adjoint fields is
\EQ{
\CW^\text{3d}_\text{adj} &\thicksim \frac{i}{R}\sum_{i,j=1}^{L}\sum_{p=1}^{n_i}\sum^{n_{j}}_{q=1}\left\{\frac{R^{2}}{4}\left[\left(\l^{(0)}_{(i-j)(p-q-1)}\right)^{2} + \left(\l^{(0)}_{(i-j-1)(p-q)}\right)^{2} - \left(\l^{(0)}_{(i-j-1)(p-q-1)}\right)^{2}\right]\right.\\
&-\text{Li}_2\left(e^{i R\l^{(0)}_{(i-j)(p-q-1)}}\right) -\text{Li}_2\left(e^{i R\l^{(0)}_{(i-j-1)(p-q)}}\right)+\text{Li}_2\left(e^{i R\l^{(0)}_{(i-j-1)(p-q-1)}}\right)\!\bigg\}\ ,\label{adj}}
where we applied the dilogarithm inversion identity $\text{Li}_{2}(e^{-iz}) + \text{Li}_{2}(e^{iz}) = \frac{1}{2}z^{2} + \pi z + \frac{\pi^{2}}{3}$ to cast the expression into a difference form. Linear and constant terms were dropped as they only yield vacuum-independent constants. Next we notice that if we add the vacuum independent constant
\eq{-\frac{iR}{4}\left(\l^{(0)}_{(i-j)(p-q)}\right)^{2}+\frac{i}{R} \text{Li}_2\left(e^{i R\l^{(0)}_{(i-j)(p-q)}}\right)} to the summand in (\ref{adj}), the quadratic terms actually become
manifestly vacuum independent and can therefore also be ignored.
Next we use the identity
\eq{\sum_{p, q=1}^{n_{i}, n_{j}} \Big[g(p-q-1) - g(p-q)\Big] = \Big(\sum_{s=-n_{j}}^{ n_{ij}-1} - \sum_{s=0}^{ n_{i}-1}\Big)g(s)\ ,} for any function $g$, to write
\eq{\CW^\text{3d}_\text{adj}\thicksim\frac{i}{R}\sum_{i,j=1}^{L}\Big(\sum_{s=-n_{j}}^{ n_{ij}-1} - \sum_{s=0}^{n_{i}-1}\Big)\bigg[\text{Li}_2\left(e^{i R\l^{(0)}_{(i-j-1)s}}\right)-\text{Li}_2\left(e^{i R\l^{(0)}_{(i-j)s}}\right)\bigg]\ .}
In this final form we can see that it is 
equal to $\CG_\text{1-loop}$ (\ref{1loop}) because contributions that depend only on one of $n_{i}$ or $n_{j}$ vanish by virtue of the Higgs branch root condition (\ref{higgs}).

The matching that we found above for a single gauge group can easily be generalized to the elliptic quiver case because the twisted superpotential is a sum over contributions from each node of the quiver.
Solving the nested Bethe ansatz (\ref{nBAE}) to leading order in the $q$ expansion as before, we find
\eq{\l^{\a(0)}_{jq} = \sum_{\b=1}^{\a-1}m_{\b}-(j-\a+1)m- q\e \mod \frac{2\pi}{R}\label{lattice1}\ ,}
for $j=1, \ldots, L$ and $q = 1, \ldots, n^{\a}_{j}$.
The on-shell value of the tree level part coincides with the classical part in the five-dimensional theory (\ref{classicalelliptic}, \ref{cselliptic}).
To compute the on-shell value of the adjoint and bi-fundamental parts (\ref{bifundamental}), it is useful to replace the term involving $\l^{\a}_{a}-\l^{\a-1}_{b}$ with $\l^{\a+1}_{a}-\l^{\a}_{b}$ (as we are summing over $\alpha$). The sum can now be performed as above and we obtain the same answer as $\CG^{\a}_\text{1-loop}$ (\ref{elliptic1loop}). 

We conclude that perturbatively $\CG = \CW^{\text{3d}}(\{n_{j}\})$ up to a vacuum-independent constant. 

\subsection{Bethe ansatz from Nekrasov's instanton partition function}

Next we demonstrate the agreement at the non-perturbative level. Our starting point is the five-dimensional generalization of the Nekrasov's instanton partition function for $\CN=(1,1)$ $SU(L)$ gauge theory with an adjoint hypermultiplet of mass $m$ \cite{Awata:2008ed}. It is written as a sum over $L$ Young tableaux $\vec Y = (Y_{1}, \ldots, Y_{L})$. The number of boxes in the $p^\text{th}$ column of the $i^\text{th}$ tableau $Y_{i}$ is denoted by $Y_{ip}$ and $|\vec Y|$ is the total number of boxes in all the tableaux. The instanton partition function, without the Chern-Simons term, is
\EQ{
&\CZ_\text{inst}(\vec{a}; \vec{Y}, m )=\\
&\sum_{\{\vec{Y}\}}q^{|\vec Y|}
\prod_{i,j=1}^{L} \prod_{p,q=1}^{\infty}\frac{\left(e^{x_{ip,jq}-m-\e}; e^{\e_2}\right)_{\infty}\left(e^{x_{ip,jq}^{(0)}-m}; e^{\e_2}\right)_{\infty}}{\left(e^{x_{ip,jq}-m}; e^{\e_2}\right)_{\infty}\left(e^{x_{ip,jq}^{(0)}-m-\e}; e^{\e_2}\right)_{\infty}}
\frac{\left(e^{x_{ip,jq}}; e^{\e_2}\right)_{\infty}\left(e^{x_{ip,jq}^{(0)}-\e}; e^{\e_2}\right)_{\infty}}{\left(e^{x_{ip,jq}-\e}; e^{\e_2}\right)_{\infty}\left(e^{x_{ip,jq}^{(0)}}; e^{\e_2}\right)_{\infty}}.
\label{Zinst5d}
}
For brevity we have set the compactification radius $R$ to unity but will restore it at the end of this section. 
Here $(f; g)_{L}=\prod_{r=0}^{L-1}(1-f g^r)$ is the Pochhammer symbol and 
\eq{
x_{ip}=a_i + p\e + Y_{ip} \e_2, \quad x_{ip}^{(0)}=a_i+p\e \qquad \text{for } i=1,\dots, L, \quad p=1,\dots, \infty,
}
while $x_{ip,jq}\equiv x_{ip}-x_{jq}$.
We can use the identify for the Pochhammer symbol 
\eq{(X ; q)_L = \frac{(X; q)_\infty}{(Xq^L; q)_\infty} \qquad \text{for } L>0,} to re-write Nekrasov's partition function (\ref{Zinst5d}) into one of  the following two equivalent forms:
\EQ{
\CZ_\text{inst} (\vec{a},\vec{Y}, m)
&=\sum_{\vec{Y}}q^{|\vec Y|}
\prod_{i,j=1}^{L}\prod_{p,q=1}^{\infty}\frac{\left(e^{x_{ip,jq}-m-\e}; e^{\e_2}\right)_{Y_{jq,ip}}}{\left(e^{x_{ip,jq}-m}; e^{\e_2}\right)_{Y_{jq,ip}}}\frac{\left(e^{x_{ip,jq}}; e^{\e_2}\right)_{Y_{jq,ip}}}{\left(e^{x_{ip,jq}-\e}; e^{\e_2}\right)_{Y_{jq,ip}}},\\
&=\sum_{\vec{Y}}q^{|\vec Y|}
\prod_{i,j=1}^{L}\prod_{p,q=1}^{\infty}\frac{\left(e^{x_{ip,jq}^{(0)}-m-\e}; e^{\e_2}\right)_{Y_{ip,jq}}}{\left(e^{x_{ip,jq}^{(0)}-m}; e^{\e_2}\right)_{Y_{ip,jq}}}\frac{\left(e^{x_{ip,jq}^{(0)}}; e^{\e_2}\right)_{Y_{ip,jq}}}{\left(e^{x_{ip,jq}^{(0)}-\e}; e^{\e_2}\right)_{Y_{ip,jq}}},
\label{trunZinst5d}
}
where $Y_{ip,jq}=Y_{ip}-Y_{jq}$. Notice that the manipulation has so far been formal in the sense that $Y_{i}$ and $Y_{j}$ can be Young tableaux containing infinite columns.
To see how the Bethe ansatz equations arise, consider the extra factor generated when one increases the length of a particular column $Y_{kr}$ of Young tableaux $Y_k$ by one unit while keeping all others fixed, i.e.~$\Delta Y_{kr} =+1$. The additional factor is
\eq{
q \prod_{j=1}^{L}\prod_{q=1}^{\infty}\frac{(1-e^{x_{kr,jq}-m})(1-e^{x_{kr,jq}+m+\e+\e_2})(1-e^{x_{kr,jq}-\e})(1-e^{x_{kr,jq}+\e_2})}{(1-e^{x_{kr,jq}-m-\e})(1-e^{x_{kr,jq}+m+\e_2})(1-e^{x_{kr,jq}})(1-e^{x_{kr,jq}+\e+\e_2})}\ . \label{extraFactor1}
}
Now if we take the Nekrasov-Shatashvili limit $\e_{2} \to 0$, then the instanton partition function becomes dominated by a saddle-point configuration which is given by the condition that the change in (\ref{trunZinst5d}) due to such increase vanishes. So setting 
\eqref{extraFactor1} to one, the saddle-point configuration can be found by solving
\eq{
1=q \prod_{j=1}^{L}\prod_{q=1}^{\infty}\frac{(1-e^{x_{kr,jq}-m})(1-e^{x_{kr,jq}+m+\e})(1-e^{x_{kr,jq}-\e})}{(1-e^{x_{kr,jq}+m})(1-e^{x_{kr,jq}-m-\e})(1-e^{x_{kr,jq}+\e})}. \label{BAE1}
}

It is important that these equations involving infinite products actually
truncate to ones involving finite products once the quantization condition \eqref{cond1} is imposed. In order to show this, we generalize the argument developed in \cite{Chen:2011sj}. To this end we define the following function:
\eq{T(x)=\frac{Q(x+\epsilon)}{Q(x)}\left[Q(x+m)Q(x-m-\epsilon)-q\frac{Q(x+m+\epsilon)Q(x-m)Q(x-\epsilon)}{Q(x+\epsilon)}\right],\label{Tx}
}
where
\eq{
Q(x)=\prod_{j=1}^L\prod_{q=1}^{\infty}\left(1-e^{x-x_{jq}}\right).
}
We see that the apparent poles of $T(x)$ coming from the zeros of $Q(x)$ in the denominator at $x=x_{jq}$ are precisely canceled by the zeros occurring at the same positions form the square bracket in the numerator, ensured by the Bethe ansatz-like equation (\ref{BAE1}). The function $T(x)$ is therefore analytic in the $x$-plane. We can further define the following functions:
\eq{
Q^{(0)}(x)=\prod_{j=1}^L\prod_{q=1}^{\infty}\left(1-e^{x-x_{jq}^{(0)}}\right), \quad P(x)=\prod_{j=1}^L(1-e^{x-a_j})
}
and the combinations
\EQ{
{\CA}(x)&=\frac{Q(x)}{Q^{(0)}(x)}, \quad \CR(x)=\frac{P(x-m-\e)P(x+m)}{P(x)P(x-\epsilon)}\ ,\\ {\cal B}(x)&=\frac{T(x)}{P(x)Q_0(x+m)Q_0(x-m-\epsilon)}\ .
}
Using these, we can rewrite the equation for $T(x)$ as
\begin{equation}\label{Tx2}
\CA(x+\epsilon)\CA(x+m)\CA(x-m-\epsilon)-{\cal B}(x)\CA(x)=q{\cal R}(x)\CA(x-\epsilon)\CA(x+m+\epsilon)\CA(x-m)\ ,
\end{equation}
From its definition $\CA(x)$ potentially has simple poles at $x=a_j+q \epsilon$, $q=1,2,\dots, \infty$ and consequently $\CA(x\pm \epsilon)$ have simple poles at $x=a_j+(q\mp 1)\epsilon$, $q=1,2, \dots, \infty$. On the other hand $\CR(x)$ on the right-hand side has zeros at $x=a_{j'}+q'\epsilon +m$, $q'=1,2,\dots, \infty$ from $P(x-m-\epsilon)$ and at $x=a_{j''}+(q''-1)\epsilon-m$, $q''=1,2,\dots, \infty$ from $P(x+m)$. 

Now suppose that the simple pole of ${\cal A}(x-\epsilon)$ on the right-hand side at
 $x=a_j+(n_j+1) \epsilon$, for some positive integer $n_j $, actually coincides with a 
zero of ${\cal R}(x)$. This requires
\EQ{\label{cond1}
&a_j+(n_j+1)\epsilon=a_{j'}+(n_{j'}+1)\epsilon + m\\
&\implies \quad a_j-a_{j'}-m=(n_{j'}-n_j)\epsilon
}
for some $j'$ and some other positive integer $n_{j'}$.\footnote{There is also another possibility obtained by choosing the zero at $x=a_{j''}+q'' \epsilon-m$ for some $j''$ and $q''$ but this gives an equivalent solution in the end and so we ignore it.} This implies that the potential pole of $\CA(x)$ on the left-hand side at $x=a_j+(n_j+1)\epsilon$ must be be removed. Then by consistency the pole of $\CA(x-\epsilon)$ at $x=a_j+(n_j+2)\epsilon$ must also not occur. Iterating this argument implies that $\CA(x)$ only has a finite set of simple poles at $x=a_j+q\epsilon$ for $q\leq n_j$. This implies that $x_{jq}=x_{jq}^{(0)}$ for $q> n_j$. In other words, the Young tableau $Y_j$ truncates at $n_j$ columns and so $Y_{jq}=0$ for $q> n_j$, provided the condition (\ref{cond1}) is obeyed. If we further choose an ordering so that $j'=j+1$ in (\ref{cond1}), we recover our quantized Higgs branch condition (\ref{quantization}) which is obviously consistent with (\ref{quantization2}). We have therefore established that the infinite products over Young tableaux (\ref{BAE1}) truncate to a finite ones if (\ref{quantization1}) is obeyed, and we recover immediately the finite set of Bethe ansatz equations discussed earlier.

Furthermore, in the limit $\epsilon_2 \to 0$, using the asymptotic limit of the non-perturbative contribution to the five-dimensional twisted superpotential obtained from the instanton partition function $\CZ_{\rm inst}(\vec{a},\vec{Y}, m)$, we find that
\eq{
\left.\CW^{\rm 5d}_{\rm inst}\right|_{a_{j} = -jm-\left(n_{j}+\frac12\right)\e}=\CY(x_{ip})-\CY(x_{ip}^{(0)}),
}
where
\EQ{
&\CY(x_{ip})=\log q\sum_{i,p} x_{ip} \\
&+\sum_{i,j}\sum_{p,q}\Big[{\rm Li}_2(e^{x_{ip,jq}-m-\epsilon})-{\rm Li}_2(e^{x_{ip,jq}-m})+{\rm Li}_2(e^{x_{ip,jq}})-{\rm Li}_2(e^{x_{ip,jq}-\epsilon})\Big]\ .
}
Using the dilogarithm inversion identity once more, it is then straightforward to 
show that this precisely coincides with the non-perturbative contributions to the three-dimensional twisted superpotential, i.e.~$\CW^{\rm 3d}(\lambda_a)-\CW^{\rm 3d}(\lambda_a^{(0)})$ where $\CW^\text{3d}$ is as given in (\ref{3dsuperpotential}). This establishes the matching of the non-perturbative parts of the five-dimensional and three-dimensional twisted superpotentials.

Let us now examine the effect of including the Chern-Simons term.
The five-dimensional Chern-Simons term contributes a constant factor in $\CZ_\text{inst}$ as \cite{Tachikawa:2004ur}
\eq{e^{-Rk \sum_{i,p}Y_{ip}\left(a_{i}+(p-\frac12)\e + \frac12Y_{ip}\e_{2}\right)}.}
The additional factor generated when we increase the $Y_{kr}$ column by one unit and take the $\e_{2} \to 0$ limit is
\eq{
e^{-Rk \left(a_{k} + \left(r-\frac12\right)\e+Y_{kr}\e_{2}\right)} = e^{-Rk (x_{kr} - \frac12\e)}\ .
}
This agrees with the contribution to the Bethe ansatz equation of a three-dimensional Chern-Simons term (\ref{bae}) once we identify the instanton positions and the Bethe roots as $\l_{ip} = x_{ip} + \e/2$.

We can also easily generalize our analysis to the five-dimensional elliptic quiver theory, whose corresponding Nekrasov partition function is given by
\EQ{
\CZ_\text{inst} (\vec{a}^{\a}, \vec{Y}^{\a}, m_\a )&=\sum_{\{\vec{Y}^{\a}\}}\prod_{\a = 1}^{K}q_{\a}^{|\vec Y^\a|}\prod_{i,j=1}^{L}
\prod_{p,q=1}^{\infty}\frac{\left(e^{x_{ip,jq}^{\a}}; q\right)_{\infty}\left(e^{x_{ip,jq}^{\a(0)}-\e}; q\right)_{\infty}}{\left(e^{x_{ip,jq}^{\a}-\e}; q\right)_{\infty}\left(e^{x_{ip,jq}^{\a(0)}}; q\right)_{\infty}}\\
&\frac{\left(e^{x_{ip}^{\a}-x^{\a+1}_{jq}-m_\a-\e}; q\right)_{\infty}\left(e^{x_{ip}^{\a(0)}-x_{jq}^{\a+1(0)}-m_\a}; q\right)_{\infty}}{\left(e^{x_{ip}^{\a}-x^{\a+1}_{jq}-m_\a}; q\right)_{\infty}\left(e^{x_{ip}^{\a(0)}-x^{\a+1 (0)}_{jq}-m_\a-\e}; q\right)_{\infty}}\ .
\label{Zinst5dQuiv}
}
where the variables are
\eq{
x_{ip}^\a =a_i^\a + p\e +Y_{ip}^\a \e_2\ , \quad x_{ip}^{\a (0)} = a_i^\a + p\e\ , \qquad p=1,\dots, \infty\ ,
}
and the Coulomb parameter $a^{\a}_i$ satisfy the twisted periodicity conditions $a_i^{\a+K} = a_i^{\a}+m$ and $a^{\a}_{i+L}=a^{\a}_i$. We can again follow exactly the same steps as in the single gauge group case and consider varying the length of the $p^\text{th}$ column of the Young tableau $Y^{\a}_{i}$. This leads to the
quantization condition
\eq{
a_i^\a - a_{i+1}^{\a+1} = -m_{\a}-\left(n_i^\a-n_{i+1}^{\a+1}\right)\e,\qquad i=1,\dots, L\ ,\label{Quantquiver}
}
and so the Young tableau $Y^\a_i$ has at most only $n_i^\a$ columns, and the summation in (\ref{Zinst5dQuiv}) is now over finite Young tableaux. The nested Bethe ansatz equation can also be recovered using the additional factors arising from varying $Y^{\a}_{ip}$ as before.

\section{Comments on Future Directions}

Using the same techniques applied in this paper, one can imagine going to one dimension higher and studying dualities between theories in six and four dimensions compactified on tori and subject to the Nekrasov-Shatashvili deformation in a two-plane. The corresponding integrable system is known as the doubly-elliptic system since in the analogue of the Ruijsenaars-Schneider model
both positions and momenta of the particles are periodic \cite{Braden:2003gv}. When there are fundamental matter fields the integrable system can be described as the anisotropic XYZ spin chain. 

There have been recent advances on computing exactly the partition function of ${\cal N}=2$ theories defined on a three-sphere \cite{Kapustin:2009kz, Jafferis:2010un, Hama:2010av}, and its ellipsoid generalization described by a squash parameter $b$ \cite{Hama:2011ea}:
\eq{b^{2} |z_{1}|^{2} + \frac{1}{b^{2}} |z_{2}|^{2} = 1, \qquad z_{1}, z_{2} \in \mathbb{C}\ .}
Using the localization technique, the partition function is reduced to a matrix integral over the real scalars $\sigma_{i}$, $i = 1, \ldots, N$ in the vector multiplet. 
\eq{
\CZ^\text{3d} = \int d^{N}\sigma~ \prod_{i=1}^{N }e^{2\pi i \zeta \sigma_{i}} \CZ^\text{3d}_\text{vector}\CZ^\text{3d}_\text{hyper}\ ,
}
Our three-dimensional theory can be thought of as a degenerate limit obtained when $b \to 0$, where the background becomes a circle of radius $R$ times a single $\Omega$-deformed plane with rotation parameter $\e$. The parameters are identified as $2\pi i b = R$ and $2\pi i b^{2} = \hbar$ (with $\e\equiv\hbar$).
The ellipsoid partition function has been shown to factorize into a vortex and an anti-vortex part \cite{Pasquetti:2011fj}. In the $b \to 0$ limit, the anti-vortex part becomes trivial and only the vortex partition function contributes in a non-trivial manner.
Indeed, the one-loop determinant from the adjoint hypermultiplets is the quantum dilogarithm $s_{b}$
\eq{\CZ^\text{3d}_\text{adj} = \prod_{i, j=1}^{N}s_{b}(\sigma_{ij} - m)\ .}
In the $b \to 0$ limit, the partition function reduces to
\eq{\CZ^\text{3d} \to \int d^{N}\sigma~\exp \left(\frac{\CW^\text{3d}}{b} \right).}
The quantum dilogarithm reduces to the classic dilogarithm and we recover the adjoint contribution to the twisted superpotential as before 
\eq{\CW^\text{3d}_\text{adj} =\frac{1}{2\pi i b}\sum_{i, j=1}^{N}\left[-\frac{(2\pi i b)^{2}(\sigma_{ij}-m)^{2}}{4} + \text{Li}_{2}(-e^{2\pi ib(\sigma_{ij}-m)})\right].}

Such three-dimensional theories on a circle have appeared repeatedly in the recent investigation of the 3d/3d duality relating three-dimensional gauge theories to theories defined on three-manifolds \cite{Terashima:2011qi, Dimofte:2011jd, Dimofte:2011ju}. The three-dimensional $\CN=4$ super Yang-Mills theory arises as the theory on the domain wall of two four-dimensional $\CN=2$ theories related by S-duality. It is identified with the $SL(2,\BR)$ Chern-Simons theory living on the cobordism of the Riemann surfaces related by the mapping class group. As such, many of the gauge theoretic quantities we have examined in this paper have nice geometric interpretations. Such three-manifold admits triangulation into hyperbolic tetrahedra with boundaries. Invariance under decomposition is translated into mirror symmetry. As the manifold is defined from gluing these tetrahedra, the gauge theory is defined by gauging the $U(1)$ flavor symmetry. The twisted superpotential matches with the hyperbolic volume of each tetrahedron and the vacuum equation is then the gluing condition between them. 

One might wonder whether the integrable structure extends to the non-degenerate squashed sphere. There are tantalizing hints from the form of the partition function, which is symmetric under exchanging $b \leftrightarrow 1/b$ due to a quantum dilogarithm property. The corresponding integrable system, if it exists, should also exhibit such modular property. The modular generalization of the XXZ spin chain has been investigated in \cite{Bytsko:2006ut}. It is natural to ask whether there is link to the squashed three-sphere. More recently, there have been attempts at the exact computation of the partition function on the five-sphere, where it is shown to reduce to integration over the instanton moduli space $\mathbb{CP}^{2}$ \cite{Kallen:2012cs, Hosomichi:2012ek}. Once computed, it should provide a lift of our five-dimensional theory on a circle.

There has been parallel development on the relation between relativistic integrable systems and dimer models \cite{Goncharov:2011hp, Eager:2011dp}. In \cite{Eager:2011dp}, it was shown that the Lax matrix of the relativistic Toda chain arises from the Kasteleyn matrix of a $Y^{p,0}$ dimer on a torus, and that the commuting Hamiltonians are generated from paths along the dimer. The $Y^{p,q}$ dimer is conjectured to give rise to a length $p$ spin chain with $p-q$ inhomogeneities. The Lax matrix of the Ruijsenaars-Schneider model involves elliptic functions of the particle position and momentum. It would be nice if this can be read off geometrically from dimer paths on a once-punctured torus.

It would be interesting to explore these directions in the future.
\acknowledgments
We thank Nick Dorey and Sungjay Lee for collaboration at various stages of this project.
HYC would like to thank the University of Cincinnati and University of Kentucky for the generous financial supports where most of this work was carried out. He is also supported in part by National Science Council and Center for Theoretical Sciences at National Taiwan University.
PZ would like to thank the Yukawa Institute for Theoretical Physics, California Institute of Technology, and the Institute for Advance Study for warm hospitality where this work was carried out, as well as the ninth Simons Workshop in Mathematics and Physics for providing a stimulating atmosphere. PZ is supported by a Dorothy Hodgkin Postgraduate Award from EPSRC and a Rouse Ball Traveling Studentship from Trinity College, Cambridge.

\end{document}